\documentclass[11pt]{article}
\usepackage{moriond,epsfig}
\def\be{\begin{equation}}
\def\ee{\end{equation}}
\def\bea{\begin{eqnarray}}
\def\eea{\end{eqnarray}}

\usepackage{amssymb}
\begin{document}


\newcommand{\abs}[1]{\vert#1\vert}
\newcommand{\fNL}{f_{\rm{NL}}}
\newcommand{\Omsd}{\Omega_{\sigma,{\rm{dec}}}}
\newcommand{\half}{\textstyle{\frac{1}{2}}}

\title{NON-GAUSSIANITY AND CONSTRAINTS FOR THE VARIANCE OF PERTURBATIONS IN THE CURVATON MODEL}

\author{ JUSSI V\"ALIVIITA$^\ast$, MISAO SASAKI$^\ddag$ and DAVID WANDS$^\ast$}

\address{$^\ast$ Institute of Cosmology and Gravitation, University of Portsmouth, Portsmouth PO1 2EG, UK\\
$^\ddag$ Yukawa Institute for Theoretical Physics, Kyoto University, Kyoto 606-8503, Japan}

\maketitle\abstracts{ Recently, the primordial non-Gaussianity in the curvaton
model has been predicted assuming sudden decay of the curvaton.
We extend the calculation to non-instantaneous decay by employing $\delta N$
formalism. 
The difference between the sudden-decay approximation and our numerical result
is larger than 1\% {\it only if} the non-linearity parameter is small, 
$-1.16<\fNL<60$. Thus it is safe to use the sudden-decay approximation when
deriving constraints for the curvaton model from WMAP3 ($\fNL<114$), but with
the Planck forecast $\abs{\fNL}<5$ one should employ the fully numerical result.
Often, the curvaton perturbations $\delta\sigma$ have been assumed to be small
compared to the background value of the curvaton field $\sigma_0$.
Consequently, the variance
$\Delta^2 =  \langle\delta\sigma^2\rangle / \sigma_0^2$ 
has been assumed to be negligible. However, the measurements of CMB or
large-scale structure perturbation amplitude do not constrain the variance if
the main contribution to it comes from the ultraviolet (UV) scales, i.e., from
smaller than observable scales.  We discuss how, even in this case, 
observational constraints on non-Gaussianity set an upper bound
to the small scale variance, $\Delta^2_{UV}<90$.}

\pagestyle{plain}
\section{Introduction}

Most inflationary models give rise to nearly
Gaussian primordial curvature perturbation. Typically, prediction for the non-linearity parameter
$\fNL$ in {\it single-field} models is of the order of $\epsilon$ --- the slow roll
parameter which must be $\lesssim 10^{-1}$ to guarantee the near
scale-invariance of the primordial perturbations.
In principle, measurement of $\fNL$ would give valuable
information on the inflaton potential, but unfortunately
such a tiny non-Gaussianity is likely to remain unobservable. 
The current
upper bound from the WMAP three-year data \cite{Spergel:2006hy} is $\abs{\fNL} < 114$
while Planck is expected to
bring this down to $\abs{\fNL} \lesssim 5$, which is still orders of magnitude larger than
the typical inflationary prediction.

Nevertheless, there are classes of {\it multi-field} models that can lead to an 
observable non-Gaussianity.
One well-motivated example is the curvaton model.\cite{LWcurvaton} 
In addition to the
inflaton $\phi$ there would be another, weakly coupled, light scalar field (e.g., MSSM flat
direction), curvaton $\sigma$, which was
completely subdominant during inflation so that the inflaton drove
the expansion of the universe. The potential could be as simple as \cite{Bartolo:2002vf}
$V=\half M^2\phi^2 + \half m^2\sigma^2$.
At Hubble exit both fields acquire some
classical perturbations that freeze in. 
However, the observed cosmic microwave (CMB) and large-scale structure (LSS)
perturbations can result from the curvaton instead of the inflaton, 
if the inflaton perturbations are much smaller than $10^{-5}$.
To simplify the analysis, in this talk \footnote{%
Jussi V\"aliviita, talk in the 40$^{\rm th}$ Rencontres de Moriond:
                  {\it Contents and Structures of the Universe},
                  La Thuile, Italy, 24/03/2006. Based on the work
                  reported in its full details in \cite{SVW}.
}, we assume that
the curvature perturbation from inflaton is completely negligible $\zeta_\phi \ll \zeta_\sigma$,
and the curvature perturbation from curvaton $\zeta_\sigma$ is such that
it leads to the observed amplitude of perturbations.

After the end of inflation the inflaton
decays into ultra-relativistic particles (``radiation'') the curvaton energy density still being subdominant.
At this stage the curvaton carries {\it pure entropy (isocurvature) perturbation} instead
of the usual adiabatic perturbation. Namely, the entropy
perturbation between radiation and curvaton is 
${\cal S}_{r\sigma} = 3(\zeta_r - \zeta_\sigma) \approx -3\zeta_\sigma$.
Since the observations have ruled out pure isocurvature primordial
perturbation\cite{Enqvist:2000hp,Enqvist:2001fu}, a mechanism --- curvaton decay into ``radiation'' 
--- that converts the isocurvature perturbation to the adiabatic one is needed
at some stage of the evolution before the primordial nucleosynthesis.

As the Hubble rate, $H$, decreases with time, eventually
$H^2 \lesssim m^2$, and the curvaton starts to oscillate about the minimum of its potential.
Then it behaves like pressureless dust (``matter'', $\rho_\sigma \propto a^{-3}$) so that its relative energy density starts
to grow with respect to radiation ($\rho_r \propto a^{-4}$). Finally, the curvaton decays into
ultra-relativistic particles leading to the standard radiation dominated adiabatic primordial
perturbations.\footnote{%
Hadn't we assumed negligible inflaton curvature perturbation,
$\zeta_\phi \approx 0$, some ``residual'' isocurvature would have
resulted if the curvaton was sub-dominant during its decay. This
would have lead to an interesting mixture of correlated adiabatic and
isocurvature perturbations which was studied in \cite{Ferrer:2004nv}.
Following the guidelines of \cite{Ferrer:2004nv} our calculation should
be straightforward to generalise. It should be noted that
{\it observations do not rule out} a correlated isocurvature component
if it is less than 20\% of the total primordial perturbation
amplitude.\cite{Kurki-Suonio:2004mn}%
} However, this mechanism may create from the initially Gaussian curvaton field
perturbation a strongly non-Gaussian primordial curvature perturbation.
The more subdominant the curvaton is during its decay
the more non-Gaussianity results in. Since the time of the decay
depends on the model parameters
(such as the curvaton mass $m$ and decay rate $\Gamma_\sigma$), the
observational upper bounds on non-Gaussianity provide a method
to constrain these parameters.
%
%
%
%

In the simplest case the (possible) non-Gaussianity results
from the second-order correction to the linear result
\begin{equation}
\zeta = \zeta_1 + \frac{1}{2}\zeta_2 =  \zeta_1 + \frac35 \fNL \zeta_1^2 \,.
\label{JVfnldef}
\end{equation}
Here $\zeta_1$ is proportional to the Gaussian field
perturbation at the Hubble exit, so it is Gaussian, and
$\zeta_1^2$ is $\chi^2$ distributed.
In this talk we briefly derive $\fNL$
in the sudden-decay approximation using a slightly different
and more general approach than in \cite{Bartolo,LR}
and then, for the first time, present the results in the case 
of non-instantaneous
decay of curvaton. 
For the full
derivation and discussion of our results see,\cite{SVW} 
where we, in addition to the second-order calculation, go to the third order,
and finally derive \emph{fully non-linear results} and \emph{full
probability density function} (pdf) of the primordial
$\zeta$ in the long-wavelength limit. 
Since, in the early universe, all today's observable
scales are super-Hubble, we take advantage of the {\it separate
universe} assumption throughout the calculations. 

\section{Sudden-decay approximation}

In the absence of interactions, fluids with a barotropic equation of
state, such as radiation ($P_r=\rho_r/3$) or the
non-relativistic curvaton ($P_\sigma=0$), have a {\it conserved} curvature
perturbation \cite{LMS}
\begin{equation}
\zeta_i(t,\vec x)
 = \delta N(t, \vec x) +
 \frac13 \int_{\bar\rho_i(t)}^{\rho_i(t, \vec x)} \frac{d\rho_i'}{\rho_i'+P_i(\rho_i')} \,,
\label{JVzetai}
\end{equation}
where $\delta N = N(t,\vec x) - \bar N(t)$ with $N(t,\vec x)$ being the
perturbed (i.e. local) number of $e$-folds of expansion until
time $t$ and $\bar N(t)$ being the average expansion. 
At the first order this
fully non-linear definition reduces to the usual definition:
$\zeta_{i 1} = -\psi_1 - H\frac{\delta_1 \rho_i}{\dot\rho_i}$ with
$\psi_1 = -\delta_1 N$.

%

Applying (\ref{JVzetai}) for the curvaton during its oscillation but before the
decay, we have
%
\begin{equation}
\label{JVzetasigma}
\zeta_\sigma(t,\vec x) = \frac13 \ln \left[ \frac{\rho_\sigma(t, \vec x)}{\bar\rho_\sigma(t)}
\right]_{\delta N=0} \,,
\end{equation}
where
$\rho_\sigma(t,\vec x) = \frac12 m^2 \sigma^2(\vec x)$
is evaluated on spatially flat ($\delta N=0$) hypersurface,
and $\sigma(\vec x)$ it the local amplitude of oscillation.
The time of the beginning of the curvaton oscillation, $t_{\rm in}$,
is defined in terms of the local Hubble rate as $H(t_{\rm in}, \vec x) = m$.
Since  $H^2 = (8\pi G/3)\rho_{\rm tot}$, the constant time $t=t_{\rm in}$  
surface is a uniform-total density hypersurface. 

In general, $\sigma_{\rm in}$ depends non-linearly on the field value
at Hubble exit $\sigma_\ast$. Thus we write\cite{LR} $\sigma_{\rm in} =
g(\sigma_\ast) = g(\bar\sigma_\ast) +
g'\delta\sigma_\ast + \half g''(\delta\sigma_\ast)^2 + \ldots$, where
$' = \partial/\partial\sigma_\ast$. (For exactly quadratic potential
$g''$ and higher derivatives vanish, but even a slight deviation from
quadratic potential can change the resulting $\fNL$ considerably \cite{nurmi} via
$g''$.)
Substituting this into (\ref{JVzetasigma}) and expanding up to second order
we obtain $\zeta_\sigma = \zeta_{\sigma1} + \half\zeta_{\sigma2} + \ldots$ with
$\zeta_{\sigma1}
 = \frac23 \frac{g'}{g}\delta\sigma_\ast$
and
$\zeta_{\sigma2}
 = - \frac32 \left( 1 - gg''/g^{\prime2} \right) \zeta_{\sigma1}^2$,
where $\delta\sigma_\ast$ is well described by a Gaussian random field
as we assume the inflaton and curvaton to be uncoupled (or only weakly
coupled), see e.g. Refs.\cite{Seery:2005gb,Lyth:2005qj}.
%
%
%
We can express this result in terms of the effective non-linearity parameter
{\it for the curvaton perturbation}, analogous to Eq.~(\ref{JVfnldef}),
$\fNL^{\sigma} = -\frac{5}{4} \left( 1 - \frac{gg''}{g^{\prime2}} \right)$.
Hence we find $\fNL=-5/4$ for the curvaton $\zeta_\sigma$ in the
absence of any non-linear evolution ($g''=0$). If the curvaton comes
to dominate the total energy density in the universe before it decays,
so that $\zeta=\zeta_\sigma$, then this is the generic prediction for
the primordial $\fNL$ in the curvaton model, as emphasised by
\cite{LR}.

Assume now that the curvaton decays instantaneously at time
$t_{\rm  dec}$ (before it has become
completely dominant) on a uniform-total density hypersurface
corresponding to $H \simeq \Gamma_\sigma$, i.e.,
when the local Hubble rate
equals the decay rate for the curvaton (assumed constant). Hence
\begin{equation}
 \label{JVbarrho}
\rho_r(t_{\rm dec},\vec x) + \rho_\sigma(t_{\rm dec},\vec x) = \bar\rho(t_{\rm dec}) \,,
\end{equation}
where we use a bar to denote the homogeneous, unperturbed quantity.
Note that from Eq.~(\ref{JVzetai}) we have $\zeta=\delta N$ on the decay
surface, and we can interpret $\zeta$ as the perturbed expansion, or
``$\delta N$''. 
%
Assuming all the curvaton decay products are
relativistic ($P=\rho/3$), we have that $\zeta$ is conserved after the curvaton
decay. 
The local curvaton and radiation densities on this decay
surface may be inhomogeneous. Indeed we have from Eq.~(\ref{JVzetai})
$\zeta_r = \zeta + \frac14 \ln \left( \rho_r/\bar\rho_r
\right)$ and 
$\zeta_\sigma = \zeta + \frac13 \ln \left( \rho_\sigma/\bar\rho_\sigma
\right)$
or, equivalently,
$\rho_r = \bar\rho_r e^{4(\zeta_r-\zeta)}$ and 
$\rho_\sigma  = \bar\rho_\sigma e^{3(\zeta_\sigma-\zeta)}$.
Requiring that the total density is uniform on the decay surface,
Eq.~(\ref{JVbarrho}), then gives the simple relation
$(1-\Omsd) e^{4(\zeta_r-\zeta)} + \Omsd e^{3(\zeta_\sigma-\zeta)} = 1 \,,$
where 
$\Omsd
= \bar\rho_\sigma/(\bar\rho_r+\bar\rho_\sigma)|_{\rm dec}$
is the dimensionless density parameter for the curvaton at the decay time.
This equation can be rewritten in the form
\begin{equation}
  e^{3\zeta_\sigma} = {\textstyle\frac{3+r}{4r}} e^{3\zeta}
  + {\textstyle\frac{3r-3}{4r}} e^{-\zeta}\,,
\label{JVnonlineq}
\end{equation}
where 
$r = \frac{3\Omsd}{4-\Omsd}$. 
Recalling Eq.~(\ref{JVzetasigma}) the LHS of
Eq.~(\ref{JVnonlineq}) is 
$\rho_\sigma(t_{\rm in},\vec
x)/\bar\rho_\sigma(t_{\rm in}) = g^2[\sigma_\ast(\vec
x)]/g^2[\bar\sigma_\ast]$.

As Eq.~(\ref{JVnonlineq}) is a fourth degree equation for $e^\zeta$,
the primordial curvature perturbation, $\zeta$, as a function
of Gaussian $\sigma_\ast$ can be solved {\it exactly}.
Remarkably, Eq.~(\ref{JVnonlineq}) was derived using fully non-linear 
definitions. Hence we have found an exact fully non-linear solution for $\zeta$
(as opposite to only second-order calculations in the literature).
Expanding this solution in $\delta\sigma_\ast$ up to second
order we then find $\fNL$, up to third order the so called 
$g_{\rm NL}$, etc. Since the exact solution is quite long,
it turns out to be easier to directly expand RHS of Eq.~(\ref{JVnonlineq})
up to any wanted order, and then equate it to the LHS, i.e., to
$g^2[\sigma_\ast(\vec x)]/g^2[\bar\sigma_\ast]$ order by order.

Up to second order the solution is
$\zeta_1 = r\zeta_{\sigma1} = r \frac23 \frac{g'}{g}\delta\sigma_\ast$
and 
$\zeta_2 = \left[ \frac{3}{2r} \left( 1 \! + \! gg''/g^{\prime2}
  \right) -2 -r \right] \zeta_1^2$.
These give the non-linearity parameter (\ref{JVfnldef}) in the sudden-decay
approximation \cite{Bartolo,LR}
\begin{equation}
\textstyle
\fNL = \frac{5}{4r}\left( 1+ \frac{gg''}{g^{\prime2}} \right) 
- \frac53 - \frac{5r}{6}\,.
\label{JVfnlfinalsd}
\end{equation}
In the limit $r\to1$, when the curvaton dominates the total energy
density before it decays, we recover the non-linearity parameter
$\fNL^\sigma$ of the curvaton
$\fNL \to \fNL^\sigma=-\frac{5}{4} \left( 1 - gg''/g^{\prime2} \right)$.
On the other hand we may get a large non-Gaussianity ($|\fNL|\gg1$) in
the limit $r\to0$, where we have
$\fNL \to \frac{5}{4r} \left( 1 + gg''/g^{\prime2} \right)$.

\section{Non-instantaneous decay}

{After the curvaton has decayed the universe is dominated by ``radiation'', with
equation of state $P=\rho/3$, and hence the curvature perturbation is
non-linearly conserved on large scales.}
Although the sudden-decay approximation gives a good intuitive
derivation of both the linear curvature perturbation and the
non-linearity parameter arising from second-order effects, it is only
approximate since it assumes the curvaton is not interacting with the
radiation, and hence $\zeta_\sigma$ remains constant on large scales,
right up until curvaton decays. In practice the curvaton energy density is
continually decaying once the curvaton begins oscillating until
finally (when $\Gamma>H$) its density becomes negligible, and during
this process $\zeta_\sigma$ does evolve\cite{GMW}. 
Another problem with results derived from the sudden-decay
approximation is that the final amplitude of the primordial curvature
perturbation, and its non-linearity, are given in terms of the density
of the curvaton at the decay time which is not simply related to the
initial curvaton density, especially as the decay time, 
$H \simeq \Gamma_\sigma$,
itself is somewhat ambiguous.

In the non-instantaneous case we can still define the first order transfer
efficiency ``$r$'' of the initial curvaton
perturbation to the output radiation perturbation, but now it
must be calculated numerically from the definition 
$r = \zeta_{1r, \rm out}/\zeta_{1\sigma, \rm in}$.
It turns out to be a function solely of the parameter \cite{GMW}
$p_{\rm in} \equiv [ \Omega_\sigma ({\frac{H}{\Gamma_\sigma}})^{1/2} ]_{\rm in}
= \frac{\bar\sigma_{\rm in}^2}{3M_{\rm Pl}^2} ( \frac{m}{\Gamma_\sigma}
)^{1/2}$, where $M_{\rm Pl}^2 \equiv 1/(8\pi G)$.

We start the numerical integration of the background Friedmann eqn together with radiation
and curvaton continuity eqns (with $\pm\Gamma_\sigma\rho_\sigma$ as a source
term) from uniform-density surface at
$t_{\rm in}$ and end it when curvaton has completely decayed
at some suitable uniform-density hypersurface $H(t_{\rm end}, \vec x) = H_{\rm
  end}$.
Then the fully non-linear primordial curvature perturbation will be
$\zeta = \zeta_{r, \rm out} = \delta N(t_{\rm end}, \vec x)$.
Repeating the calculation with different initial values $\sigma_{\rm in}$
we find the function $N(\sigma_{\rm in})$.
From this we then calculate the first and second order quantities
by employing $\delta N = N'\delta\sigma_\ast + \half N''(\delta\sigma_\ast)^2 +
\ldots \equiv \zeta_1 + \half \zeta_2 + \ldots$, 
and finally \cite{LR} $\fNL = \frac{5}{6} N''/(N')^2$.
From Fig.~1(a) we see that if $\fNL>60$ ($r<0.02$) or $\fNL<-1.16$ ($r>0.95$), the
sudden-decay result differs from the non-instantaneous decay result
less than $1\%$.%
\footnote{After this talk the
non-instantaneous decay calculation was done in \cite{Malik:2006pm} employing
{\it second order perturbation equations} 
instead of $\delta N$ formalism which we use. The results
agree after taking into account that the authors of \cite{Malik:2006pm}
compare $\fNL$s resulting from the fixed $p_{\rm in}$ while
we prefer to compare $\fNL$ when the sudden decay and non-instantaneous decay
produce same $r$, or, in other words, {\it the same
observable first order curvature perturbation}. To achieve this, we need to
start from different $p_{\rm in}$ for sudden decay than for non-instantaneous
decay. The non-trivial mapping between $p_{\rm in}$ and $r$
is demonstrated in Fig.~1(b)
for the non-instantaneous decay. 
}
In the both cases, as $r\to 1$ we have $\fNL \to -\frac{5}{4}$.
As $r \to 0$, the sudden-decay result is
$\fNL \to \frac{5}{4} \frac{1}{r} ( 1 + g g''/{g'}^2 ) - 1.67$, whereas
the non-instantaneous decay gives $\fNL \to \frac{5}{4} \frac{1}{r} ( 1 + g g''/{g'}^2 ) - 2.27$.
We can write the full result in a form
$\fNL = \frac{5}{4} \frac{1}{r} ( 1 + g g''/{g'}^2 ) +\frac{5}{4} h(r)/r^2$,
where the function $h(r)$ is defined by equation $r' = [2r + h(r)]g'/g$, and determined
numerically in the non-instantaneous case.
For the sudden decay we have from (\ref{JVfnlfinalsd}) 
$h(r) =  -\frac{4}{3}r^2 -\frac{2}{3}r^3$.

\begin{figure}
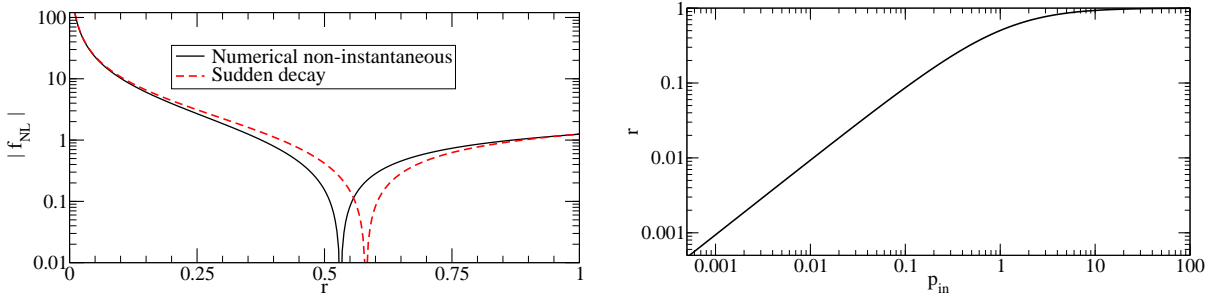

\begin{center}
\includegraphics[width=0.48\textwidth]{fNLfigure}
\hspace{0.02\textwidth}
\includegraphics[width=0.48\textwidth]{rvspMoriondArxiv}
\end{center}
\caption{(a) The non-linearity parameter $\fNL$ as a function of curvature
perturbation transfer efficiency $r = \zeta_{r, \rm out} / \zeta_{\sigma, \rm in}$.
The analytical approximative, i.e., sudden-decay result ({\it red dashed line})
crosses zero at $r = 0.58$ and is negative for $r>0.58$. The exact numerical
result ({\it black solid line}) crosses zero at $r=0.53$. 
(b) $r$ vs $p_{\rm in}$.
}
\end{figure}

\section{Large variance on small scales}

Finally, we consider the possibility of a large
small-scale variance\cite{LM06}
$\Delta^2 \equiv \langle (\frac{\delta\sigma_{\rm in}}{\bar\sigma_{\rm in}})^2 \rangle$. 
Let us name the observable CMB scales as infrared ($IR$) and smaller 
scales as ultraviolet ($UV$) so that
$\lambda_{IR} \gg \lambda_{UV} \gg H_{\rm in}^{-1}$. As the
observations require $\langle \zeta_1^2 \rangle_{IR} \lesssim 10^{-9}$, they
set constraint for $\frac{4}{9} r^2 \Delta^2_{IR}$.
However, observations do not directly constrain $\Delta^2_{UV}$.
Allowing large $\Delta^2_{UV}$ we find
$\fNL = \frac{5}{4}  (1+\Delta^2_{UV}) \frac{1}{r}
( 1 + g g''/{g'}^2 )
+\frac{5}{4} h(r)/r^2$, where $h(r)$ remains same as without large
variance. Thus, large \emph{homogeneous} small-scale variance modifies the first term of $\fNL$
only. Recalling that $r\le 1$, and $-54 < \fNL < 114$ (from WMAP3 \cite{Spergel:2006hy}), we
find an upper bound  $1+\Delta^2_{UV} < \frac{4}{5}\times 114 = 91$.

\end{document}